\documentclass[letter]{aa} % for the letters 
\usepackage{graphicx}
\usepackage{txfonts}
\usepackage{natbib}
\begin{document}
   \title{Detection of an asymmetry in the envelope of the carbon Mira R~Fornacis using VLTI/MIDI
\thanks{Based on observations made with ESO telescopes at La Silla
Paranal Observatory under program IDs 080.D-0231 and 084.D0361.}}
   %\subtitle{I. Overviewing the $\kappa$-mechanism}

   \author{C. Paladini
          \inst{1}
          \and
          S. Sacuto\inst{2}
          \and
          D. Klotz\inst{1}
          \and
          K. Ohnaka\inst{3}
          \and
          M. Wittkowski\inst{4}
          \and
          W. Nowotny\inst{1}
          \and
          A. Jorissen\inst{5}
          \and
          J. Hron\inst{1}
          %\and
          %K. Eriksson\inst{2}
          %\fnmsep\thanks{Just to show the usage
          %of the elements in the author field}
          }

   \institute{University of Vienna, Dept. of Astrophysics,
              T\"urkenschanzstrasse 17, A-1180 Vienna, Austria\\
              \email{claudia.paladini@univie.ac.at}
              \and
		  Department of Physics and Astronomy, Uppsala University, Box 516, 75120 Uppsala, Sweden
              \and
              Max-Planck-Institut f\"ur Radioastronomie, 53121 Bonn, Germany
              \and
              ESO, Karl-Schwarzschild-Str. 2, 85748 Garching bei M\"unchen, Germany
              \and
                            Institut d'Astronomie et d'Astrophysique, 
Universit\'e Libre de Bruxelles, CP 226, Boulevard du Triomphe, 1050, Bruxelles, Belgium\\
             }

   \date{Received; accepted}

% \abstract{}{}{}{}{} 
% 5 {} token are mandatory
 
  \abstract
   {}
   { We present a study of the envelope morphology of the carbon Mira R~For with VLTI/MIDI. 
     This object is one of the few asymptotic giant branch (AGB) stars that underwent a dust-obscuration event. 
     The cause of such events is still a matter of discussion. 
     Several symmetric and asymmetric scenarios have been suggested in the literature.}
   {Mid-infrared interferometric observations were obtained separated by two years. The observations probe different depths 
    of the atmosphere and cover different pulsation phases. The visibilities and the differential phases were interpreted using GEM-FIND, a tool for fitting spectrally dispersed interferometric observations with the help of wavelength-dependent geometric models.}
   {We report the detection of an asymmetric structure revealed through the MIDI differential phase. This asymmetry is observed at the same baseline 
     and position angle two years later. %This is the first asymmetry detected in the mid-infrared for a carbon-rich Mira. 
     The observations are best simulated with a model that includes a uniform-disc plus a Gaussian envelope plus a point-source. 
     The geometric model can reproduce both the visibilities and the differential phase signatures.}
   {Our MIDI data favour explanations of the R For obscuration event that are based on an asymmetric geometry.
     We clearly detect a photocentre shift between the star and the strongly resolved dust component. This might be caused by a
     dust clump or a substellar companion. However, the available observations do not allow us to distinguish between the two options.
     The finding has strong implications for future studies of the geometry of the envelope of AGB stars:
     if this is a binary, are all AGB stars that show an obscuration event binaries as well? Or are we looking at 
     asymmetric mass-loss processes (i.e. dusty clumps) in the inner part of a carbon-rich Mira?}

   \titlerunning{Detection of an asymmetry in the envelope of the carbon Mira R~Fornacis}
   \authorrunning{Paladini et al.}
   \maketitle
%
%________________________________________________________________

 \section{Introduction}

The C-star \object{R For} is classified as a Mira variable in the General Catalogue of Variable Stars \citep{sam09}.
Objects of this variability class show photometric variations with amplitudes larger than 2.5~mag in the $V$ band, 
and with long periods ($>100$ days). The light curves of Mira variables are usually described as 
regular with sinusoidal behaviour. Nevertheless, long-term observations of these objects show erratic drops in their brightness \citep[][and references therein]{bar96}.

The variability of R~For was discussed for the first time by \cite{fea84}, and later by \cite{leb88}. Both works
concluded that the light curve showed a deep minimum in 1983.
This decrease was attributed to an increased dust obscuration, the reason for which is (still) unknown.

Four scenarios were envisaged by \cite{fea84} to explain the obscuration event, the first two scenarios
were symmetric, and the latter two involved an asymmetric geometry of the envelope. The scenarios involve:
\begin{enumerate}
\item the ejection of a spherical shell by the star;
\item enhanced dust-condensation in an already existing spherical shell; 
\item an asymmetric dust ejection in a preferential direction (i.e. a disc, or a more complex structure); 
%\cite{whi97} hypothesised that the obscuration event is due to 
%an eclipse caused by a dust cloud ejected along the line of sight. 
\item an asymmetric dust ejection in a random direction. 
\end{enumerate}
%The drop in the light curve is due to an eclipse by the dust cloud ejected along the line of sight in cases 3 and 4. 
\cite{leb88} and \cite{win94} supported the first two scenarios, which reproduced 
the photometrical data with dynamic models in spherical symmetry. 
Advanced 2D and 3D model atmospheres by \cite{woi05} and \cite{fre08} predicted 
the formation of concentrated dust clouds in the inner part of the atmospheres, 
and a spherical distribution on a large scale. Such a morphology might explain the dust obscuration events. 
However, up to now those models were not compared with observations.
On the other hand, these obscuration events have similarities with those observed for R~Coronae Borealis stars {\citep[R~CrB;][]{cla12}}. 
In the latter case, the third and the fourth scenario would be more likely explanations \citep{bri11, jef12}.
\cite{lea07} used VLTI/MIDI to study the circumstellar environment of the R~CrB star RY~Sgr. 
This star showed a drop in the $J$ and $H$ light curve similar to the one observed for R~For.
The observations of RY~Sgr showed that the star is surrounded by a single dusty cloud located at $\sim$100 stellar radii ($\sim$30 AU).
\cite{whi97} suggested a binary-related effect as an alternative explanation. Exceptionally deep minima in the light curves 
of single O-rich Miras are not observed, but they are fairly common for symbiotic Miras. 

This letter presents high angular resolution observations in the mid-infrared that 
allow us to probe the dust-forming region of the nearby AGB star R~For.
Observations close in time, with different position angles and baseline lengths,
the differential phase information, and the comparison with geometric models can help to distinguish 
possible deviations from a spherical structure. 
This allows one to confirm or reject the different scenarios suggested in the past.
Observations and data reduction procedures are presented in Sect.~\ref{obs.sect}.
The study of the morphology by means of geometric models is described in
Sect.~\ref{morphstu.sect}, and is discussed in Sect.~\ref{disc.sect}.
%**********************uvcov.fig 1*******************************************************
\onlfig{1}{
\begin{figure}[!thtp]
\includegraphics[angle=0.,width=.5\textwidth]{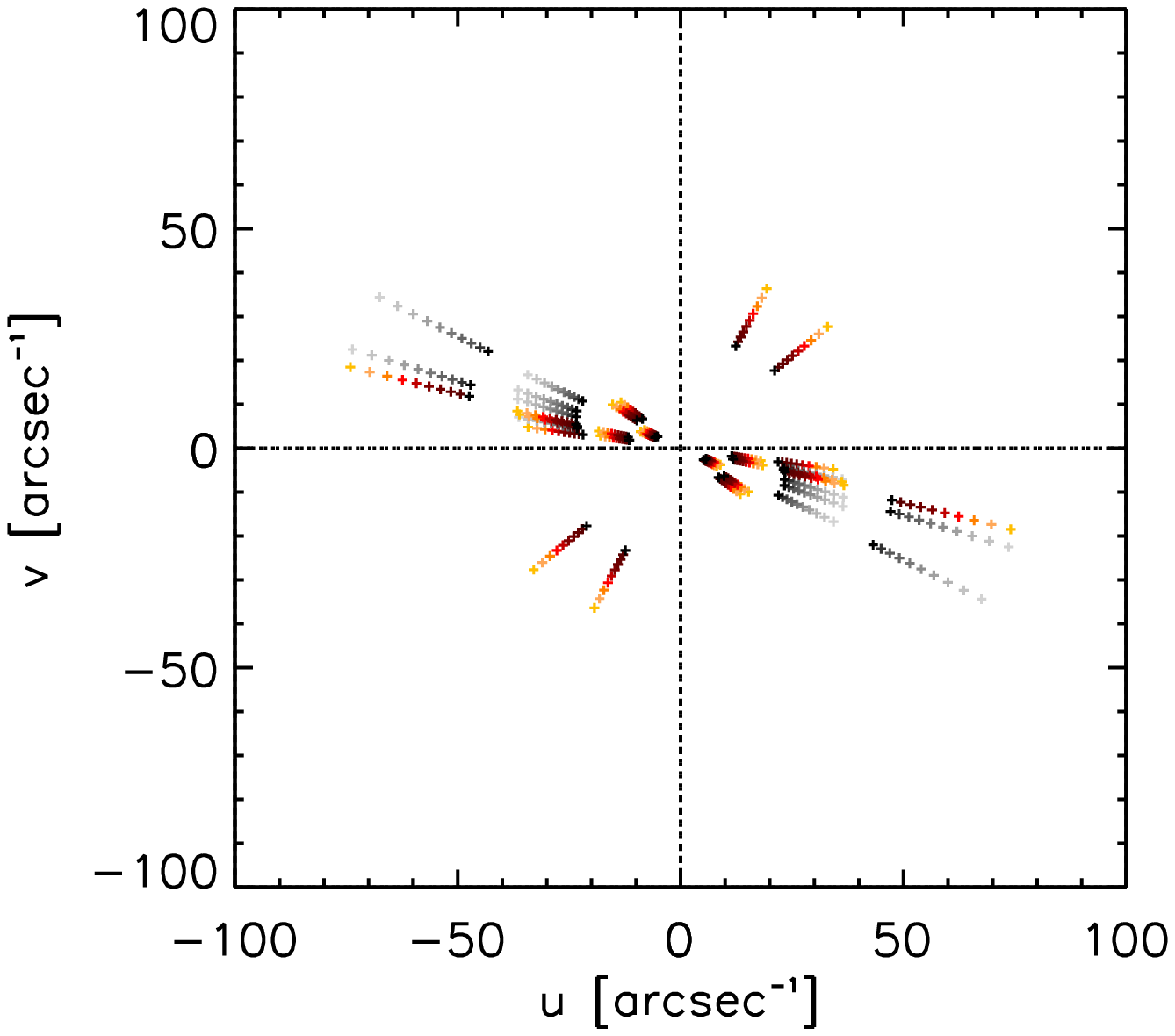}
\caption{\label{uvcov.fig}\small{$uv$ coverage dispersed in wavelength of the observations 
acquired during the programs {\sc080.D-0231} (grey-scale), and {\sc084.D-0361} (red-scale).  
The colour code ranges from 8-12.5~$\mu$m, with light colours indicating shorter wavelengths.
The sample is restricted to the observations used for this work (marked with (b) and (d) in Table~\ref{obs.tab}).}} 
   \end{figure}
   }
%**********************************************************************************************
%__________________________________________________________________
\section{Observations and data reduction}
\label{obs.sect}
To study the morphology of the close circumstellar environment of R~For, 
high angular resolution measurements were acquired in the mid-infrared with the ESO-VLTI/MIDI facility (program IDs 080.D-0231 and 084.D0361, Ohnaka and Sacuto respectively). 
In Table~\ref{obs.tab} the 29 visibility points observed for the science target are presented. 
%**********************obs.tab 1; online*******************************************************
\onltab{1}{
\begin{table*}[h!]
       \centering
         \caption{\label{obs.tab}\small{Journal of the MIDI observations for the C-star R~For.}}
       \smallskip
\begin{small}
         \begin{tabular}{l l l l l l l}
         \hline
         \hline
         \# & Date \& Time         & Phase\tablefootmark{*} & VLTI Configuration & $B_{\rm{p}}$   & PA    & Seeing     \\
            &    [UT]              &       &                    & [m]           & [deg] & [arcsec]\\
         \hline
          01\tablefootmark{a} & 2007-10-04 03:17:13  & 1.07 (a) & G0-H0              & 26          & 43  & 0.82 \\
          02\tablefootmark{a} & 2007-10-05 04:58:33  & 1.08 (a) & G0-H0              & 31          & 61  & 0.88 \\
          03\tablefootmark{a} & 2007-10-29 03:17:27  & 1.14 (a)  & A0-K0              & 122         & 60  & 1.13 \\
          04 & 2007-12-02 01:47:31  & 1.23 (b) & D0-H0              & 63          & 66  & 2.05 \\
          05 & 2007-12-02 02:51:00  & 1.23 (b) & D0-H0              & 63          & 73  & -   \\
          06\tablefootmark{a} & 2007-12-02 03:00:50  & 1.23 (b) & D0-H0              & 63          & 74  & -  \\
          
          07 & 2007-12-06 01:10:17  & 1.24 (b) & A0-K0              & 125         & 63  & 1.35 \\
          08 & 2007-12-06 02:36:01  & 1.24 (b) & A0-K0              & 127         & 73  & 0.88 \\
          09 & 2007-12-10 00:58:46  & 1.25 (b) & D0-H0              & 63          & 64  & 1.17 \\
         10 & 2007-12-10 01:50:54  & 1.25 (b) & D0-H0              & 64          & 70  & 1.28 \\
         11 & 2007-12-12 03:11:41  & 1.25 (b) & D0-H0              & 61          & 79  & 1.44 \\
         12  & 2007-12-29 01:54:08 & 1.30 (b) & D0-H0              & 61          & 78  & 1.32 \\
         13  & 2008-02-27 01:07:55 & 1.45 (c) & D0-H0              & 67          & 99  & 0.99 \\
         \hline
         14\tablefootmark{a} & 2009-09-06 07:32:39 &  2.89 (d) & D0-H0 & 63  & 67 & 1.08  \\ 
         15 & 2009-09-06 09:06:52 &  2.89 (d) & D0-H0 & 62  & 77 & 0.95 \\ 
         16 & 2009-09-07 05:43:04 &  2.89 (d) & G0-H0 & 28  & 52 & 0.76 \\ 
         17 & 2009-09-07 06:18:38 &  2.89 (d) & G0-H0 & 30  & 57 & 0.48  \\ 
         18 & 2009-09-07 09:10:11 &  2.89 (d) & H0-K0 & 31  & 78 & 1.30 \\ 
         19 & 2009-09-07 09:44:55 &  2.89 (d) & H0-K0 & 30  & 81 & 1.45 \\ 
         20 & 2009-09-21 07:53:15 &  2.93 (d) & A0-G1  & 126& 76  & 1.08 \\
         21 & 2009-09-26 09:00:58 &  2.94 (d) & D0-G1  & 68 & 152 & 0.75 \\ 
         22 & 2009-09-27 06:01:02 &  2.94 (d) & D0-G1  & 71 & 130 & 0.68 \\
         23 & 2009-09-27 08:43:21 &  2.94 (d) & D0-H0 & 57  & 82 & 0.65 \\ 
         24 & 2009-10-01 05:21:32 &  2.95 (d) & E0-G0   & 15& 63 & 1.15 \\ 
         25 & 2009-10-01 05:59:46 &  2.95 (d) & E0-G0   & 16& 67 & 1.09 \\
         26 & 2009-11-18 03:31:27 &  3.08 (a$ ^\prime$) & A0-G1  & 128& 72  & 0.90 \\
         27 & 2009-11-18 03:50:10 &  3.08 (a$ ^\prime$) & A0-G1  & 127& 74  & 0.65 \\ 
         28 & 2010-01-11 03:13:02 &  3.22 (b$ ^\prime$) & D0-H0   & 38& 96 & 0.65 \\ 
         29 & 2010-01-11 03:48:17 &  3.22 (b$ ^\prime$) & D0-H0   & 45& 92 & 0.76 \\
         \hline 
              \end{tabular}
              \tablefoot{
              \tablefoottext{*}{Estimated from AAVSO light curve ({\tt{http://www.aavso.org/aavso-research-portal}}).
			\tablefoottext{a}{Data excluded from astrophysical interpretation, 
			because they were not fulfilling one or more of the five criteria for the quality assessment listed in Sect.~2.2 of \cite{klo12a}.}
              }
              }
\end{small}
        \end{table*}
        }% end of onltab
        %**********************************************************************************************
Twenty-four points out of the original sample turned out to be of good quality.
In this work we concentrate on the data marked in Table~\ref{obs.tab} with (b) and (d): 
these observations were obtained (within each group) at roughly the same pulsation phase 
($\pm~0.03$), but they sample different spatial scales of the object.
Therefore, these two groups are ideal for studying the morphology of the star. 
The remaining observations will be presented in a forthcoming publication (Paladini et al., in prep.), 
where we compare them with dynamic model atmospheres \citep{mat10}.

The $uv$ coverage dispersed in wavelength used for this work is plotted in Fig.~\ref{uvcov.fig}.
Every observation of R~For is paired with the observation of at least one calibrator. 
The list of calibrators used for the data reduction is given in Table~\ref{calib.tab}.
%*******************************calib.tab 2; online***************************************************************
\onltab{2}{
\begin{table}[]
  \caption{\label{calib.tab}\small{List of the calibrators.}}
  \begin{tabular}{l l l}
    \hline
    \hline
    Calibrator & $F_{12}$ & $\vartheta_{\rm{UD}}$  \\
    &[Jy]   &  [mas] \\
    \hline
         HD~224935          &86.90  & $7.24 \pm 0.03 $~\tablefootmark{a} \\
         HD~25025           &109.6  & $8.76 \pm 0.624$~\tablefootmark{b} \\ %Gam eri
         HD~20720           &112.8  & $10.03\pm 0.1  $~\tablefootmark{c} \\
         HD~12929           &77.80  & $7.43 \pm 0.52 $~\tablefootmark{a} \\ %alf ari                            
         HD~48915           &143.1  & $5.85 \pm 0.15 $~\tablefootmark{a} \\ %alf ari                            
         \hline
  \end{tabular}
  \tablefoot{The $12~\mu$m flux listed in col.~2 is the IRAS flux. 
    \tablefoottext{a}{ESO/MIDI database.}
    \tablefoottext{b}{JMMC database.}
    \tablefoottext{c}{\cite{dyc96}.}%should be 98
  }
\end{table} 
}%end onlintab{2}
%**********************************************************************************************
The data were reduced with the standard data reduction pipelines MIDI Interactive Analysis ({\textsc MIA}), 
and Expert Workstation ({\textsc EWS}) package version 1.7.1 \citep{koe05, jaf04}. 
A detailed description of the data quality check and error determination is given in \cite{klo12a}.

\begin{figure*}[thp]
\includegraphics*[angle=0,width=0.9\textwidth,bb=70 371 483 625]{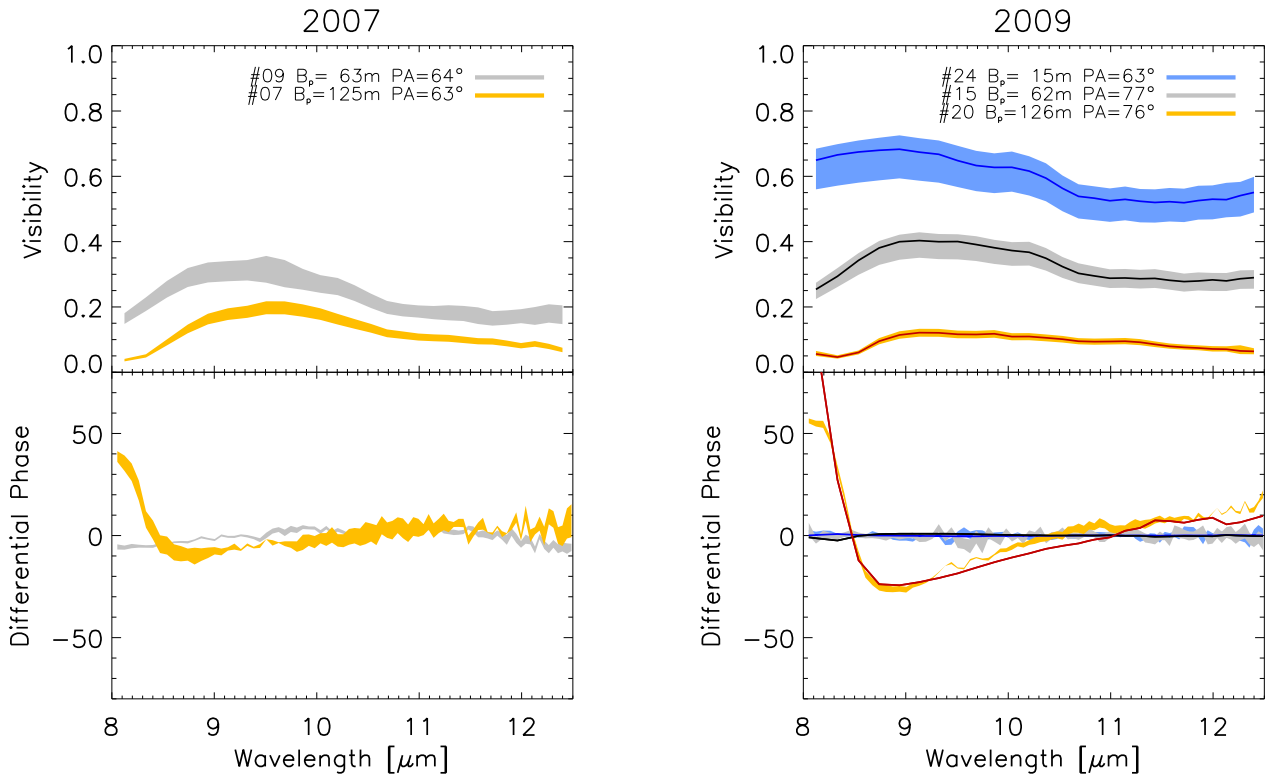}
 \caption{\label{vardiffpha.fig}\small{
The shaded-region show the calibrated visibilities (upper panels) and differential phases (lower panels) versus wavelength 
for the selected observations of 2007 (left) and 2009 (right). 
The overplotted lines show the best-fitting model (details in Sect.~\ref{morphstu.sect}).}}
\end{figure*}

The derived visibilities (upper panels of Fig.~\ref{vardiffpha.fig}) exhibit the typical $tilde-$shape 
of $N-$band interferometric observations of C-stars with a SiC feature present at 
$11.3~\mu$m \citep{ohn07, sac11}. We recognise a drop in the $8-9\,\mu$m region, 
where the main contributors to the opacity are C$_2$H$_2$ and HCN (Fig.~\ref{wave-flux.fig}). 
We note that the long baseline visibilities of 2009 are systematically lower 
than the observations of 2007, indicating a more extended dust shell in 2009.
%************************VARIAB.fig & DIFFPHA.FIG 2
%**********************************************
The differential phase is a crucial observable for this work because it carries information about the morphology of the source. 
This corresponds to the difference in phase between the different spectral
channels. The EWS software performs a linear fit to the index of water vapour refraction to clean the phase signal from the
atmospheric effects \citep{jaf04,tub04}.   
In R~For all differential phases show zero signal, except for those observed with the longest baselines (see lower panels of 
Fig.~\ref{vardiffpha.fig}).
The phase information in these observations clearly points to a non-central symmetric structure.
Signatures in the differential phase have been reported in the literature for two silicate J-type C-stars: IRAS~18006-3213 by \cite{der07}, 
and BM~Gem by \cite{ohn08}. These objects show silicate emission that probably originates 
from a circum-binary or a circum-companion disc. 
Many C-stars are known to show photospheric asymmetries \citep{rag06,cru12}, while very few extreme objects depart from spherical morphology in the inner dust shell \citep[e.g. IRC+10216 and CIT6; ][]{mac80, wei98, mon00, tut00, cha07}. To our knowledge 
\emph{this is the first time that MIDI has detected a signature of deviation from central symmetry for a carbon Mira variable
through the differential phase}.

%***************************************MORPHSTU.SECT***************************************************************
\section{Morphological interpretation}
\label{morphstu.sect}
The wavelength-dependent geometrical model-fitting tool GEM-FIND \citep{klo12b} was used to model the 
interferometric observations obtained at comparable visual phases in 2009 ((d) in Table~\ref{obs.tab}). 
GEM-FIND is based on the Levenberg-Marquardt least-squares minimisation method and allows the fitting of different centro-symmetric 
and asymmetric geometric models to the dispersed calibrated visibilities. 
The fitting is limited to the wavelength range $<\,12.5~\mu$m
to avoid any bias that may be caused by the noise that usually affects MIDI measurements at longer wavelengths.
The best-fitting model visibilities were then used to 
derive synthetic differential phases, which were compared to the observed ones. 

Spherical and aspherical optically thick (i.e one-component)  
and optically thin (i.e., two-component) models were tested. The  
best-fitting model (Fig.~\ref{model.fig}) consists of (i) a uniform disc (UD) that represents the central star,  
and (ii) an extended centrosymmetric Gaussian distribution that represents the  
envelope. 
%*************************************model.fig 3; online*********************************************************
\onlfig{3}{
\begin{figure}[thp]
\includegraphics[angle=0,width=0.45\textwidth]{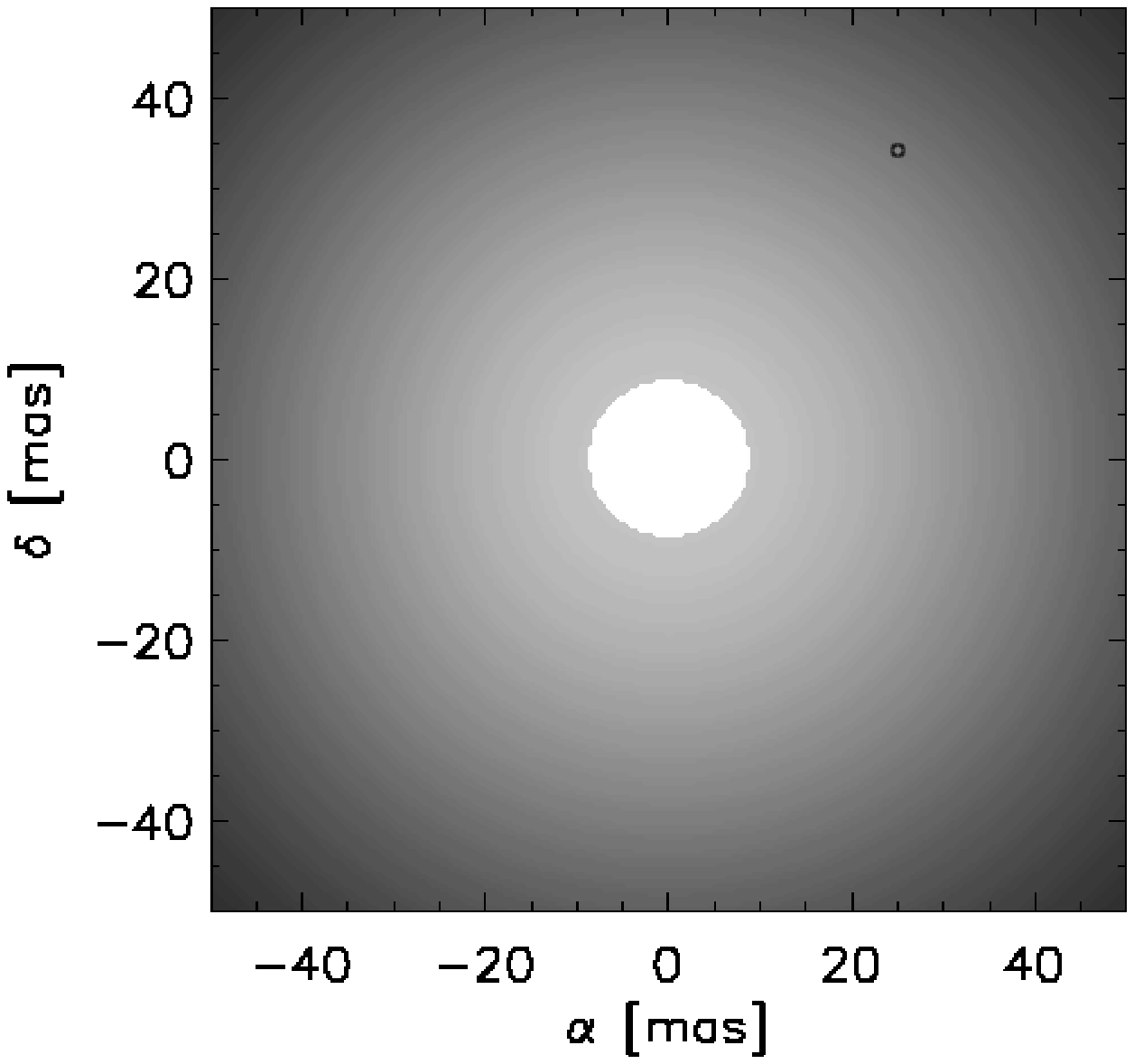}
\caption{\label{model.fig}\small{Normalised intensity distribution at 10~$\mu$m 
in the sky plane of the geometric model that best fits the R~For MIDI data of 2009.
The position of the asymmetric structure is indicated with a circle.
The axis orientation is chosen according to the standard with north up and east left.}}
\end{figure}
}
%**********************************************************************************************
Furthermore, to account for the non-zero differential phases at long baselines, (iii) a third component corresponding to a  
Dirac delta function was added. This geometric model is defined by six parameters listed in 
Table~\ref{param.tab} with the resulting best-fitting values (or range of values for wavelength-dependency).
Because the angular information of the 2007 observations is limited, we were not able to apply the same 2D model fitting 
strategy as above to those data (see Sect.~\ref{disc.sect}).
%****************************param.tab 3; online******************************************************************
\onltab{3}{
\begin{table}[]
\begin{center}
\begin{footnotesize}
\caption{\label{param.tab}Parametric description of the geometrical
model CircGauss+UD+Dirac.}
\begin{tabular}{ll}
\hline\hline
Parameters & Results\\
\hline
$\theta_{\mathrm{star}}$, mas& 15-22 \\ 
FWHM, mas      & 70-130 \\
$F_{\rm{env}}/F_{\rm{star}}$ &  0.75-1.08 \\
$F_{\rm{blob}}/F_{\rm{star}}$ &  0.02-0.07\\      
$\rho$,  mas   & 42 \\
$\phi$   &  - 36$^\circ$\\    
\hline
\end{tabular}
\end{footnotesize}
\end{center}
\footnotesize{
{\textbf{Notes.}} 
A range of values is listed for the wavelength dependent parameters. 
$\theta_{cen}$\ldots diameter of central star;
FWHM\ldots full width at half maximum of the Gaussian envelope; 
$F_i/F_j$\ldots flux ratios for Gaussian envelope/central star and blob/central star;
$\rho$, $\phi$\ldots separation and position angle of the component from the primary star. Detailed description in \cite{klo12b}.}
\end{table}
} 
%**********************************************************************************************
Observed and modelled visibilities and differential 
phases are displayed in the right panel of Fig.\,\ref{vardiffpha.fig}. 
We note a very good agreement between model and observations for the visibilities and for the differential phases. 
The model flux ratios of the Gaussian envelope and of the blob over the central star 
are plotted in Fig.\,\ref{wave-flux.fig}. 
For comparison we show the contributions of relevant molecular and 
dust species that characterise the spectrum of a C-rich AGB star at these wavelengths \citep{now11}. 
The flux ratio of the blob over the central star follows the shape of the
C$_2$H$_2$ contribution, with a slight shift of the minimum towards the shorter wavelengths (where the HCN contribution has a minimum).
A weak drop can be recognised around 11.3~$\mu$m where the SiC feature is located. The spectral feature in the differential phase
(Fig.~\ref{vardiffpha.fig}) and the flux ratio (Fig.\,\ref{wave-flux.fig}) suggest that the asymmetry originates in the C$_2$H$_2$+HCN layer
(8-9~$\mu$m region), and possibly also in the SiC-forming region. 
The features of $F_{\rm{blob}}/F_{\rm{star}}$ could be caused by the star or by the nature of the blob. This will be discussed in Sect.~\ref{disc.sect}.\\
The best-fitting model (reduced $\chi^2=0.29$, 4 free parameters) yields a FWHM for the Gaussian envelope of 70-130 mas. 
These values are consistent with single-dish telescope mid-infrared
observations by \cite{lag11} that showed an unresolved
source with an angular resolution of $\sim 0.3$ arcsec. The model also yields a separation of 
the point source of 42\,mas at an angle of -36$^\circ$.
We derived a $K-$band diameter of 8 mas (unpublished AMBER observations, PI Ohnaka), 
i.e. the asymmetric structure is located at about 5 stellar radii 
\citep[$\sim$30 AU assuming 680 pc as distance;][]{whi06} inside the envelope of the star. 
The position of the Delta function has to be treated with caution, because it is derived by fitting data with limited angular coverage. The fit 
presented here provides one example of a quantitative solution, but more complex morphologies cannot be excluded.

\section{Discussion}\label{disc.sect}
As already mentioned, a differential phase signature, similar to that observed for R~For, was detected 
for two J-type carbon stars. IRAS~1800-3213, presented by \cite{der07}, 
shows a signature in the differential phase that is in the same wavelength region (i.e. $8-9\,\mu$m) as the one observed for R~For.
\cite{der07} mentioned that long baselines probe the part in the atmosphere where dust emission is strongly (spatially) resolved. The authors speculated that the differential phase signature is caused by an offset of the photocentre between the strongly resolved dust region and the unresolved stellar photosphere. This might be explained by a circumbinary disc.
The second object, BM~Gem, presented by \cite{ohn08}, shows a non-zero differential phase signature between $9-11\,\mu$m. 
The authors interpreted this as due to the presence of a circum-companion disc.

The detected asymmetry for R~For excludes the symmetric scenarios 1. and 2. hypothesised by \cite{fea84}.
The plausible explanations are 
(i) a feature intrinsic to the star, such as a dust blob that moves through the circumstellar environment of R\,For; or 
(ii) an unresolved companion that is embedded in the circumstellar envelope.

The dust blob might have been ejected in a preferential direction or randomly, as stated in scenarios 3 and 4.
The expansion velocity of the outer envelope of R~For is 16.9~km~s$^{-1}$ \citep{gro99}. Using the determined separation of 42 mas, 
and assuming that the blob moved with a constant velocity of 5\,km\,s$^{-1}$ over the last 30 years,
\citep[a reasonable value for the inner envelope, predicted by dynamic model computations;][]{now10}, 
it is worthwhile to mention that the event of 1983 would correspond to the moment at which such a 
structure was formed near the surface of the star. 
Recent results from investigations of R~CrB stars \citep{jef12} indicate the presence of dust clouds with different grain-sizes. Nevertheless, \cite{bri11} did not detect any differential phase signature for these objects, arguing that the signatures of dust clumps
are not strong enough to induce significant spectral signatures like the one we observed.
This might not be the case for AGB stars because R~CrB stars are much hotter and H-deficient (i.e. no C$_2$H$_2$), 
therefore carbon nucleation proceeds differently \citep{goe92}.
IRC+10216 could be a very interesting test-case to check if the clumps can produce a phase signature in carbon stars.
Indeed, several dust clumps were detected in the atmosphere of this star \citep{han98, wei98, wei02, lea06} at spatial scales comparable to the one we obtained with the GEM-FIND modelling ($\sim 30$~AU).

\begin{figure}[thp]
\includegraphics*[angle=0,width=0.45\textwidth,bb=62 532 467 795]{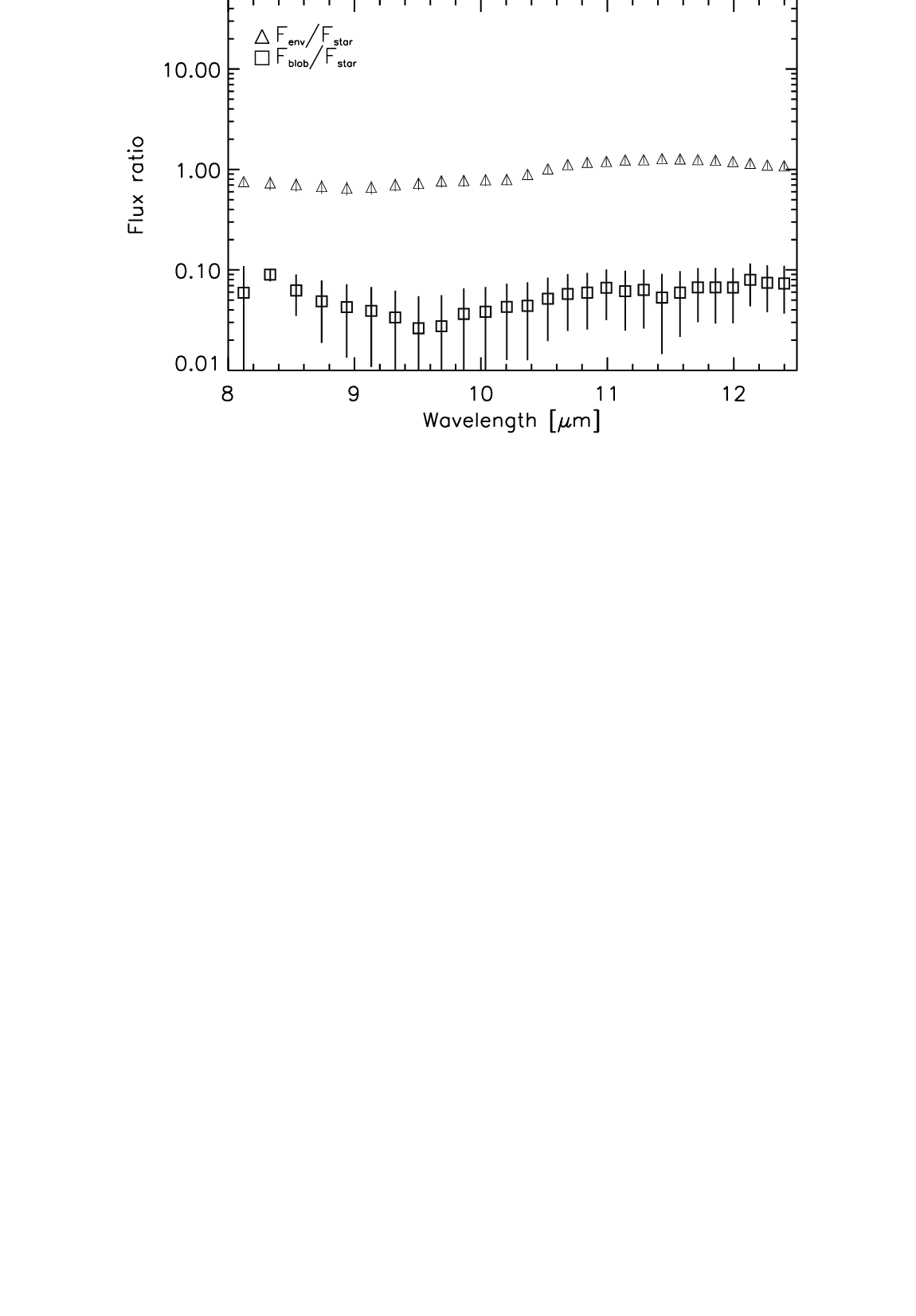}
\includegraphics*[angle=0,width=0.45\textwidth,bb=62 397 467 795]{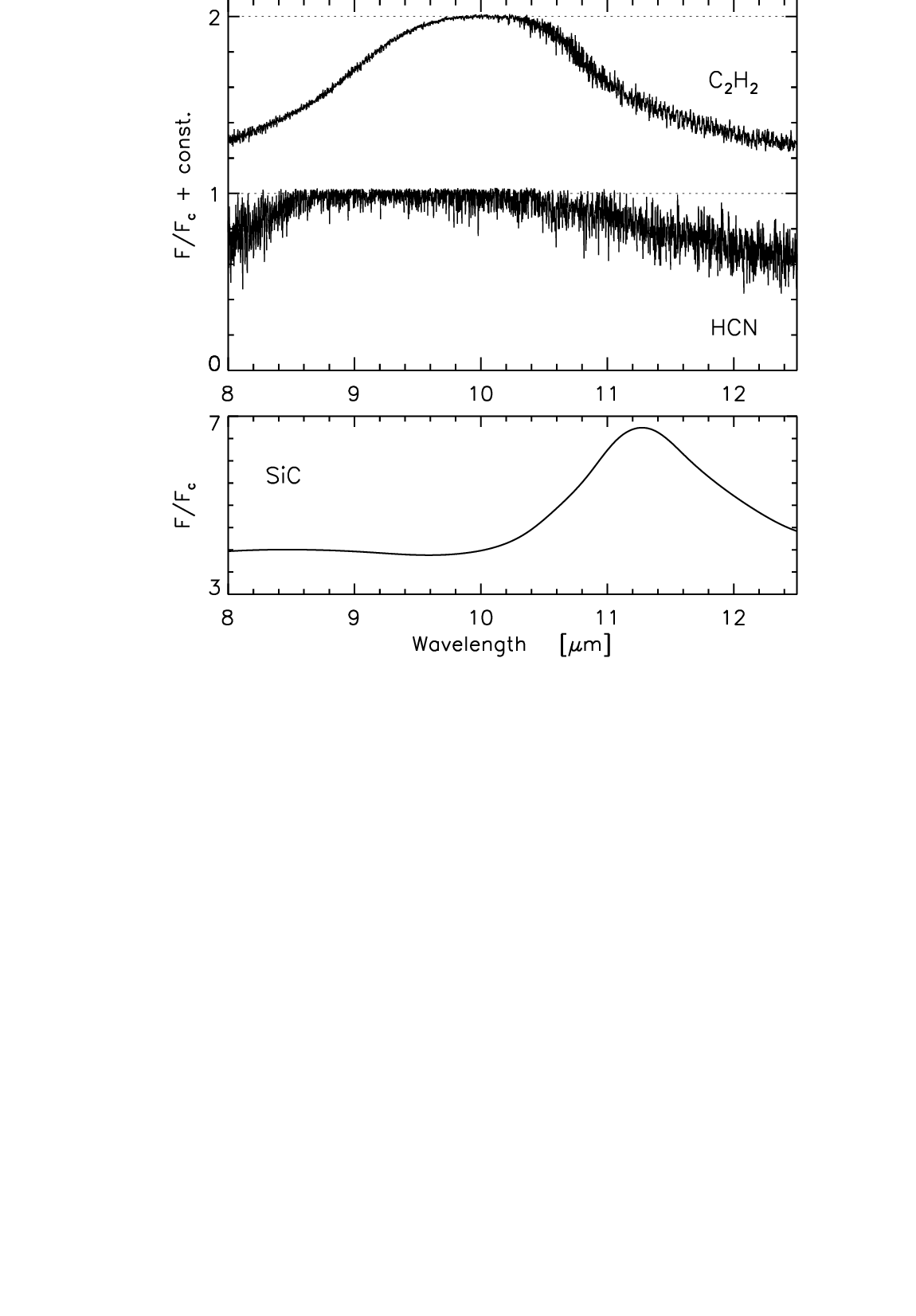}
 \caption{\label{wave-flux.fig}\small{Upper panel: Wavelength-dispersed flux ratio values of the best-fitting models. 
Other panels: Contributions by different sources in typical 
spectra of C-rich AGB stars for the MIDI wavelength range. The synthetic 
spectra are plotted with continuum-normalised fluxes $F/F_{\rm c}$, for 
detailed descriptions see \cite{ari09} or  Nowotny et al. 
(2010). The middle panel illustrates the individual molecular absorption 
due to C$_2$H$_2$ and HCN with the help of a hydrostatic model 
atmosphere, for details we refer to Fig.\,4 of \cite{ari09}. 
The lower panel shows the characteristic emission due to circumstellar SiC 
dust on the basis of a dynamic model atmosphere \citep[e.g.][]{hoe03,now10}, 
computed following the approach of \cite{sac11} to artificially impose SiC grains. This dust species was detected
in the ISO spectrum of R~For \citep{cle03}. }}
\end{figure}

The binary option is quite unrealistic for a stellar companion for the following reasons.
Archive GALEX observations do not show any evidence of UV excess, which automatically excludes the presence of a warm white dwarf. 
By translating the flux ratio (see Fig.~\ref{wave-flux.fig}) into magnitudes, this would imply that the companion has a brightness of $\sim2.2$\,mag at 12~$\mu$m. A faint giant star is very unlikely because such an object would be resolved. 
A main sequence star also turns out to be too massive and thus would disrupt (because of mass accretion) 
the spherical symmetry detected at low spatial frequencies for the envelope. 
We cannot exclude that the companion is a substellar object orbiting the envelope.
There are suggestions in literature that such a situation might be common among AGB stars \citep{sos07}.
%Strong mass accretion is not expected, although the planet would carve a gap in the envelope, but would in principle not destroy
%the symmetry of the circumstellar environment. 
In this case, only low-mass accretion is expected \citep{lec07}, leaving a signature only at the high spatial frequencies
probed by long baselines.

The differential phase signature was detected with the same baselines at the same position angles after two years.
Assuming that the flux ratio of the point source over the central star 
is constant between 2007 and 2009 and that the position of the companion 
as well as the dust blob would change from 2007 to 2009, the following scenarios are possible: 
(i) for $1~M_{\sun}$ central star and a Jupiter mass companion (10$^{-3}\,M_\odot$) the orbital period of the companion would be 185\,yr and would result in a change of the position angle of $\pm$4\,$^\circ$; 
(ii) assuming a dust blob that is moving outwards at a uniform velocity of 5\,km\,s$^{-1}$ would result in a change of 
the separation of 2.4\,mas. 
In both cases the asymmetry would be still observable within the two years with the chosen MIDI configuration. 
Therefore, the available observations do not allow us to reject any of these two asymmetric scenarios.

This letter presents one interpretation for the detected asymmetry in the dust envelope of R~For.
More complex morphologies cannot be excluded a priori because of the limited coverage of the available observations.
Interferometric imaging with the second-generation instrument VLTI/MATISSE \citep{lop06} and a detailed polarimetric investigation 
are needed to clarify the nature of the detected structure.

%************************wave-flux-ratio.FIG 4**
%**********************************************
\begin{acknowledgements}
This work was supported by the Austrian Science Fund FWF 
under project number AP2300621 and P21988-N16. We thank the referee E. Lagadec 
for useful comments, and O. Chesneau, B. Aringer, A. Chiavassa for fruitful discussions.

\end{acknowledgements}

%************************OBS.TAB*************************************************
 %************************CALIB.TAB**********************************************************************
%**********************************PARAM.TAB*********************************************************
\end{document}